# Anderson localization of light with topological dislocations

Valery E. Lobanov,[1] Yaroslav V. Kartashov,[1,2] Victor A. Vysloukh,[3] and Lluis Torner[1]

[1]*ICFO-Institut de Ciencies Fotoniques, and Universitat Politecnica de Catalunya, 08860 Castelldefels (Barcelona), Spain*
[2]*Institute of Spectroscopy, Russian Academy of Sciences, Troitsk, Moscow Region, 142190, Russia*
[3]*Departamento de Fisica y Matematicas, Universidad de las Americas—Puebla, 72820 Puebla, Mexico*

We predict Anderson localization of light with nested screw topological dislocations propagating in disordered two-dimensional arrays of hollow waveguides illuminated by vortex beams. The phenomenon manifests itself in the statistical presence of topological dislocations in ensemble-averaged output distributions accompanying standard disorder-induced localization of light spots. Remarkably, screw dislocations are captured by the light spots despite the fast and irregular transverse displacements and topological charge flipping undertaken by the dislocations due to the disorder. The statistical averaged modulus of the output local topological charge depends on the initial vorticity carried by the beam.

The phenomenon of localization of electronic wave functions in the presence of a static disordered potential, nowadays known as Anderson localization, was predicted in solid-state physics by P. W. Anderson more than 50 years ago [1,2]. After the discovery of such effect it was realized that localization is a general wave phenomenon that may occur for waves of different physical nature. Anderson localization has been experimentally observed in various areas of science [3], including microwaves [4,5], acoustics [6,7], matter waves in Bose-Einstein condensates [8,9], and optics [10-16]. A particularly important manifestation of the effect is the so-called transverse Anderson localization [13], which occurs when the wavepacket remains localized in the transverse spatial axes, but not along the evolution or propagation coordinate.

Optical systems provide a unique landscape to explore such phenomena [14]. Current technology allows fabrication of photonic lattices with periodic shallow transverse modulations of the refractive index, where controllable disorder invariable in the direction of light beam propagation, can be introduced [17-19]. Transverse Anderson localization has been observed in optically-induced [18] and in fabricated lattices [19-24], in both two-dimensional [18,22,23] and one-dimensional [19-21,24] geometries. The impact of the inhomogeneity of disorder [25], potential gradients [24], nonlinearity [18,19,26,27], and interfaces [20,23,27,28] have been analyzed. The possibility of Anderson localization in lattices with Bragg mechanism of light guiding [29] and in PT-symmetric structures [30] has been also addressed. In most theoretical and experimental studies performed to date Anderson localization has been studied for the case of bell-shaped, Gaussian-like input beams, matching the lowest-order fundamental modes of the isolated channel forming the lattice. Rare exceptions include investigations of the effect of a transverse phase modulation on the evolution of broad one-dimensional inputs in disordered discrete lattices [31]. However, Anderson localization of excitations carrying topologically nontrivial wavefronts has never been addressed to date. The problem is relevant to several areas of physics where Anderson localization occurs (in particular, in condensed matter, superfluid phenomena, superconductivity, or quantum atomic system trapped in optical lattices, to name just a few) and where robust topological dislocations appear nested in smooth wavefronts. A salient example of such a robust topological dislocation is an optical vortex [32]. Vortices are ubiquitous entities in Nature, thus their study is relevant to many areas of physics. In optics, they occur as screw phase-dislocation nested around a point where the field is undefined, thus the light intensity must vanish. Since the screw phase dislocation results from an interference effect, the dynamical evolution of the vortex core is strongly affected by the presence of external potentials, anisotropies, asymmetries or astigmatism. Such dynamics directly impacts the evolution of the value of the topological charge carried by the beam.

In this paper we address the dynamics of evolution of the topological charge of screw phase dislocations that results from the propagation of optical vortices in disordered optical lattices under conditions where transverse Anderson localization may take place. We find that a weak diagonal disorder introduced into a periodic lattice of hollow-core shallow waveguides illuminated by a light beam carrying an optical vortex results not only in localization of the optical fields, but also in the presence of topological dislocations in the corresponding light spots, in spite of the highly irregular dynamics, that includes fast and continuous charge flipping, undertaken by the wavefront dislocations. Namely, we found that the statistically averaged modulus of the topological charge of the beam in the central region of the pattern remains close to the input value at all propagation distances.

We consider linearly polarized light beams propagating along the $\xi$-axis of a disordered array of parallel ring-like, hollow-core waveguides. In the frame of the paraxial wave approximation that describes the propagation of monochromatic light beams with a width $w \gg \lambda$ in media with a shallow refractive index modulation, the evolution of the dimensionless amplitude of the light field $q$ is governed by the Schrödinger equation (for a derivation see, e.g., [33,34]):

$$i\frac{\partial q}{\partial \xi} = -\frac{1}{2}\left(\frac{\partial^2 q}{\partial \eta^2} + \frac{\partial^2 q}{\partial \zeta^2}\right) - R(\eta,\zeta)q. \qquad (1)$$

The transverse coordinates $\eta, \zeta$ and propagation distance $\xi$ are normalized to the characteristic transverse scale $r_0$ and the diffraction length $k_0 r_0^2$, respectively. The refractive index $R(\eta,\zeta) = \sum_{n=-\infty}^{+\infty}\sum_{k=-\infty}^{+\infty} p_{n,k} G(\eta-\eta_n, \zeta-\zeta_k)$ in the waveguide array represents a superposition of the hollow-core channels $G(\eta,\zeta) = \exp[-(r-\rho)^2/w^2]$ with width $w$, radius $\rho$, and guiding parameters $p_{n,k}$, placed in the nodes $\eta_n = nd$, $\zeta_k = kd$ of a square grid. Here $r = (\eta^2 + \zeta^2)^{1/2}$ and $d$ is the spacing between neighboring hollow-core waveguides. We address lattices with diagonal disorder, where the depths of the individual waveguides $p_{n,k} = p_0(1+\delta_{n,k})$ are randomized, with $\delta_{n,k}$ being a random refractive index perturbation uniformly distributed within the interval $[-p_d; p_d]$, and $p_0$ is the mean refractive index. We select the parameters of the array in accordance with experimentally accessible values [20,22] and set $p_0 = 6$, $\rho = 1.2$, $d = 4$, $w = 0.5$, which correspond to a refractive index contrast $\sim 7\times 10^{-4}$, waveguide width $\sim 5\,\mu\text{m}$ and period $\sim 40\,\mu\text{m}$. A propagation distance $\xi = 1$ at the wavelength $\lambda = 632$ nm corresponds to a propagation length of some $\sim 1.44$ mm. The hollow-core shallow waveguide arrays considered here can be fabricated using existing direct laser-writing technologies [20], or induced optically [18] by arrays of mutually-incoherent higher-order Bessel beams [35,36] or by specially designed non-diffracting patterns such as those reported in [37].

In all cases we use as the input beam a vortex-carrying eigenmode of an isolated hollow-core waveguide, whose shape $q|_{\xi=0} = W_m(r)\exp(im\phi)$ can be obtained directly from Eq. (1), where the function $W_m(r)$ describes radial field distribution of stationary mode, $\phi$ is the azimuthal angle, $m$ is the winding number or topological charge of the mode, and angular field variation is given by the $\exp(im\phi)$ term. In the isolated waveguide such a vortex-carrying mode propagates without distortion. However, the picture changes dramatically in the waveguide array, where overlap of the modal fields of neighboring waveguides may lead to light tunneling from the excited waveguide into nearest neighbors. We found that the evolution of such modes even in a regular waveguide array depends on the value of the topological charge. Figure 1 shows initial intensity distributions and diffraction patterns at relatively short distances $\xi$ for various values of $m$. When $m$ increases the input eigenmode

becomes less localized and its overlap with neighboring waveguides increases leading to stronger diffraction. Note the pronounced cross-like intensity distributions, indicating the enhanced diffraction in the directions parallel to the $\eta,\zeta$ axes for the $m=1$ beam, which is in obvious contrast to the diagonal expansion obtained for the $m=0$ and $m=2$ beams. Thus, one expects that anisotropic diffraction may notably affect the localization process in the presence of disorder. It should be stressed that the topology of the input beam strongly affects the symmetry of the modes excited in the far-away waveguides when the excitation reaches them in the process of diffraction. For $m=1,2$ the modes with dipole and quadrupole symmetries are predominantly excited, while for $m=0$ only axially-symmetric modes are visible.

In order to quantify the localization of light in the studied setting, we generated $Q \sim 10^3$ realizations of the disordered arrays for each level of disorder $p_d$ and each topological charge considered. Using a split-step Fourier method, we integrated Eq. (1) up to large distances $\xi = L > 10^3$ for each disorder realization, using a vortex mode of the isolated waveguide with topological charge $m$ as an input. For the statistical analysis the output intensity distributions $|q_i(\eta,\zeta)|^2$ for each disordered array were averaged over the ensemble of array realizations and the averaged integral form-factor $\chi_{av}$ was calculated as:

$$I_{av}(\eta,\zeta) = Q^{-1} \sum_{i=1}^{Q} |q_i(\eta,\zeta,\xi)|^2 \big|_{\xi=L},$$
$$\chi_{av}(\xi) = Q^{-1} U^{-2} \sum_{i=1}^{Q} \iint |q_i(\eta,\zeta,\xi)|^4 \, d\eta d\zeta, \qquad (2)$$

where $U = \iint |q_i(\eta,\zeta)|^2 \, d\eta d\zeta$ is the beam power. The inverse integral form-factor $\chi_{av}^{-1/2}(\xi)$ characterizes the width of the localized core of the averaged intensity distribution, with only a minor contribution from the low-intensity background, which is discriminated due to the fourth power of the field in the integrand in Eq. (2). The smaller the value of $\chi_{av}^{-1/2}(\xi)$, the higher the localization.

Figure 2 illustrates the ensemble-averaged output intensity distributions obtained for input light with various topological charges and for increasing disorder levels $p_d$. While in each realization the output intensity distribution may be strongly asymmetric, in the averaged picture one obtains patterns with well-defined symmetry. Anderson localization occurs because even weak disorder qualitatively changes the eigenmodes of the array – in contrast to periodic Bloch waves in regular array the eigenmodes of two-dimensional disordered array are all localized. In the case of diagonal disorder the Anderson-localized eigenmodes are centered on the waveguides with highest refractive index. The characteristic scale of the modes is dictated by the disorder level: at small $p_d \sim 0.01$ the modes extend over tens of waveguides, while at $p_d \sim 0.2$ they concentrate nearly on a single waveguide. Narrow input vortices excite a set of eigenmodes with different weights given by the overlap integrals between mode shapes and input field. The subsequent evolution is dictated by beating between excited eigenmodes that have different propagation constants. At high disorder levels $p_d \sim 0.2$ only a limited set of well-localized eigenmodes is excited, because projections on modes residing far from the excited waveguide are negligible. In this case the input beam experiences almost no spreading upon evolution (see Fig. 2, last column). In contrast, at small disorder a number of poorly localized modes are excited and the initially localized vortex-carrying beam considerably spreads across the array (Fig 2, first column).

The dependences of the statistically averaged output inverse form-factor on the disorder level $p_d$ are presented in Fig. 3, for $m=1,2$. The degree of localization of the intensity distributions of vortex modes monotonically grows with increasing disorder level. To confirm that Anderson localization was reached for all disorder levels considered, in Fig. 4(a) we plot the evolution of the aver-aged inverse form-factor as a function of the distance $\xi$. The plot shows the clear transition from the initial ballistic spreading to the localization regime. The localization length [the distance at which $\chi_{av}^{-1/2}(\xi)$ reaches its steady-state value] rapidly diminishes with increasing $p_d$. In Fig. 4(b) we show the cross-section of the averaged output intensity distribution at $\eta = 0$ for the input beam with $m = 1$ in logarithmic scale. The nearly perfect triangular profile indicates that localized modes feature exponentially decaying tails. A similar picture was obtained for $m = 2$ input beams.

The central result of this Letter is that when the input beam carries a vortex with a single-charge ($m = 1$) dislocation, *in spite of the high refractive-index anisotropy (or, more properly, transverse inhomogeneity)* introduced by the disorder, our numerical experiments predict that at almost any propagation distance and disorder realization, *one encounters a phase dislocation within the central channel of the array*. The dislocation may have charge $m = 1$ or $m = -1$. Similarly, when the input beam carries a double-charge vortex ($m = 2$), one finds that the central channel may contain no dislocations, one dislocation or, very often, two single-charge dislocations. To study the statistical behavior of the topological charge of the dislocation, we monitored both, the averaged topological charge $m_{av}$ and the averaged modulus of topological charge $|m|_{av}$. Data was obtained by calculating for each disorder realization the circulation of the local phase $\phi_i$ along a closed contour surrounding the central point, given by

$$m_{av}(\xi) = (2\pi Q)^{-1} \sum_{i=1}^{Q} \oint d\phi_i,$$
$$|m(\xi)|_{av} = (2\pi Q)^{-1} \sum_{i=1}^{Q} |\oint d\phi_i|. \tag{3}$$

We found that the averaged modulus of the topological charge remains nearly constant upon evolution and that it is only slightly smaller than the modulus of input charge $m$, as shown in Fig. 4(c). For example, at $p_d = 0.15$ one gets $|m|_{av} \approx 0.9$ when the input beam carries a vortex with $m = 1$ and $|m|_{av} \approx 1.65$ when the input beam carries a double-charge vortex ($m = 2$). Note that the statistically averaged values $|m|_{av}$ depend on the particular values of the parameters used on the calculations, because they are affected by the precise irregular dynamics followed by the dislocations during their evolution. The important point is that, in spite of such rich and complex dynamics followed by the wavefront dislocations during the light evolution in the presence of the high transverse refractive-index inhomogeneity introduced by the disorder, the local charge in the central waveguide of the array is correlated to the topology of the input excitation.

In all cases that we analyzed, the statistically averaged charge $m_{av}$ was found to be close to zero. This is due to the occurrence of irregular dynamical topological charge inversions. Individually, such charge inversions are reminiscent of the charge flipping that occurs in simpler but similar physical settings, such as astigmatic optics [38], deformed optical fibers [39], or periodic photonic lattices [40]. In the disordered lattices, one observes that, because of the local astigmatism caused by the presence of the disorder, the input dislocation moves towards the regions where the field intensity vanishes where, e.g., it may annihilate with an antivortex from one of the vortex-antivortex pairs born in that region, while the dislocation with the opposite charge moves toward the center of the waveguide. Note that, conceptually, charge inversion may occur at the transverse infinity, which acts as a reservoir of topological charges [38].

The charge inversion visible in Fig. 5, which shows the phase and intensity distributions inside the array for a particular disorder realization, can also be interpreted as the dynamical interference of the modes of the perturbed waveguides. Thus, in the isolated radially-symmetric ring waveguide the guided modes appear in degenerated doublets $h_m(r,\phi)\exp(ib\xi) = w(r)\exp(\pm im\phi + ib\xi)$ with equal propagation constants $b$, but opposite charges. The presence of azimuthal modulations of the refrac-tive index results in lifting of this degeneracy and splitting of each doublet into two modes $s_m \exp(ib_s\xi)$, $a_m \exp(ib_a\xi)$, which for weak refractive index deformations represent symmetric $s_m = (h_m + h_m^*)/2$ and antisymmetric $a_m = (h_m - h_m^*)/2$ combinations of degenerated states with slightly different propagation constants $b_{s,a} = b + \delta b_{s,a}$. The evolution of an input beam with topological charge $m$ that effectively excites $s_m$ and $a_m$ modes of the deformed waveguide is described by the beating pattern $C s_m \exp(ib_s\xi) + Q a_m \exp(ib_a\xi)$, where $C, Q$ are the projections of the input beam on the $s_m, a_m$ modes. Propagation over the distance $\pi/|b_s - b_a|$ results in the accumulation of a $\pi$ phase difference between the $s_m$ and $a_m$ components and thus in the inversion of the topological charge even in a single deformed waveguide.

This scenario becomes more complex when the refractive index modulation is induced by the second adjacent waveguide because of the light-tunneling phenomenon. In this case charge inversion in the excited waveguide occurs simultaneously with power beating between adjacent guides and both these processes feature exactly the same periodicity. In the disordered lattice, where selected channel is surrounded by eight neighbors with different guiding parameters, the power beating between multiple channels and charge inversion becomes irregular [Fig. 5(a)]. However, one still can clearly distinguish the characteristic stages when the ring-like pattern with phase dislocation in the central waveguide first transforms into the dipole one, and then transforms back into the ring structure [Fig. 5(c)]. As shown in Fig. 5(a), due to replacement of dislocations by their oppositely charged counterparts in the central channel, the topological charge oscillates between two dominant values $m = \pm 1$, nearly instantaneously passing the $m = 0$ stage. This is the reason why $|m|_{av}$ takes on the value that is only slightly below the input charge $m$.

Summarizing, we addressed the study of the propagation of light in disordered arrays of hollow-core waveguides illuminated by beams carrying optical vortices. We found that Anderson localization of the ring-shaped light spots is accompanied by the dynamic, statistical localization of screw topological dislocations nested in the spots. In contrast to simple naive expectations, the dislocations do not fade away and disappear in the transverse plane as a consequence of the increasing disorder, thus yielding a standard Anderson localization of light spots with no topological charge. Rather, the topological wavefront dislocations undergo highly-complex transverse dynamics, including anisotropy-induced charge flipping, but on average a dislocation with topological charge with a modulus close to the input value is encountered within the Anderson-localized spot. The results reported here are relevant also for other areas of physics where the phenomenon of Anderson localization takes place, including the evolution of matter waves in disordered optical potentials.

# Figure captions

Figure 1.  (Color online) Intensity distributions at different distances showing discrete diffraction in a regular array for the input beam with $m=0$ (a), $m=1$ (b), and $m=2$ (c). White circles in (a) indicate positions of waveguides in the array. All quantities are plotted in arbitrary dimensionless units.

Figure 2.  (Color online) Averaged output intensity distributions at $\xi=1500$ for different disorder levels and input beams with $m=0$ (a), $m=1$ (b), and $m=2$ (c). White circles in the last panel in row (c) show positions of waveguides in the array. All quantities are plotted in arbitrary, dimensionless units. All quantities are plotted in arbitrary dimensionless units.

Figure 3.  Averaged output inverse form-factor at $\xi=1500$ versus disorder level for the input beams with $m=1$ (a) and $m=2$ (b). All quantities are plotted in arbitrary dimensionless units.

Figure 4.  (a) Averaged inverse form-factor versus propagation distance at $p_d=0.08$ (curve 1) and $p_d=0.175$ (curve 2) for the input beam with topological charge $m=1$. (b) The cross-section of the averaged output intensity distribution at $\eta=0$ for the input beam with $m=1$ drawn in the logarithmic scale. The disorder level is $p_d=0.15$. (c) Averaged modulus of topological charge of the central phase singularity versus $\xi$ at $p_d=0.15$. All quantities are plotted in arbitrary dimensionless units.

Figure 5.  (Color online) (a) Topological charge of the wavefront singularity located in the central waveguide versus propagation distance. Field modulus (b) and phase (c) distributions corresponding to the points marked by circles in panel (a). In all cases the input beam carries topological charge $m=1$ and disorder level is $p_d=0.15$. The topological charge at a given distance $\xi$ is defined using phase circulation over closed contour indicated in (b) and (c). All quantities are plotted in arbitrary dimensionless units.

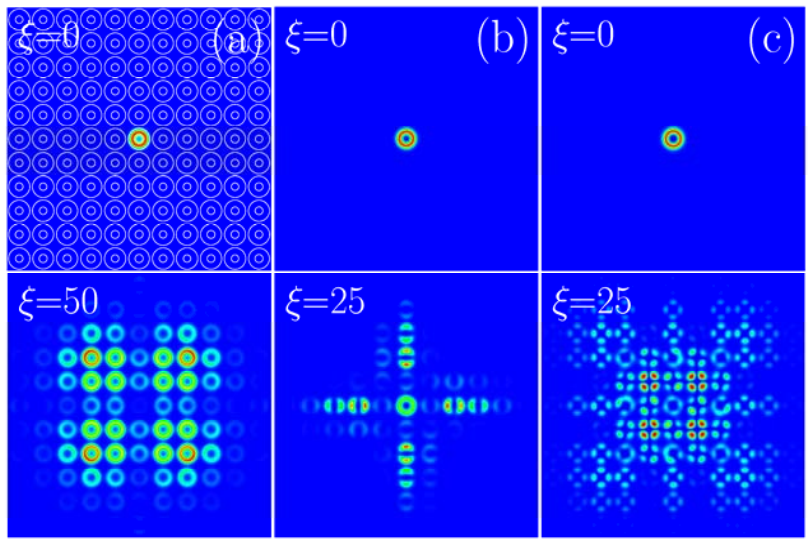

Figure 1. (Color online). Intensity distributions at different distances showing discrete diffraction in a regular array for the input beam with $m=0$ (a), $m=1$ (b), and $m=2$ (c). White circles in (a) indicate positions of waveguides in the array. All quantities are plotted in arbitrary, dimensionless units.

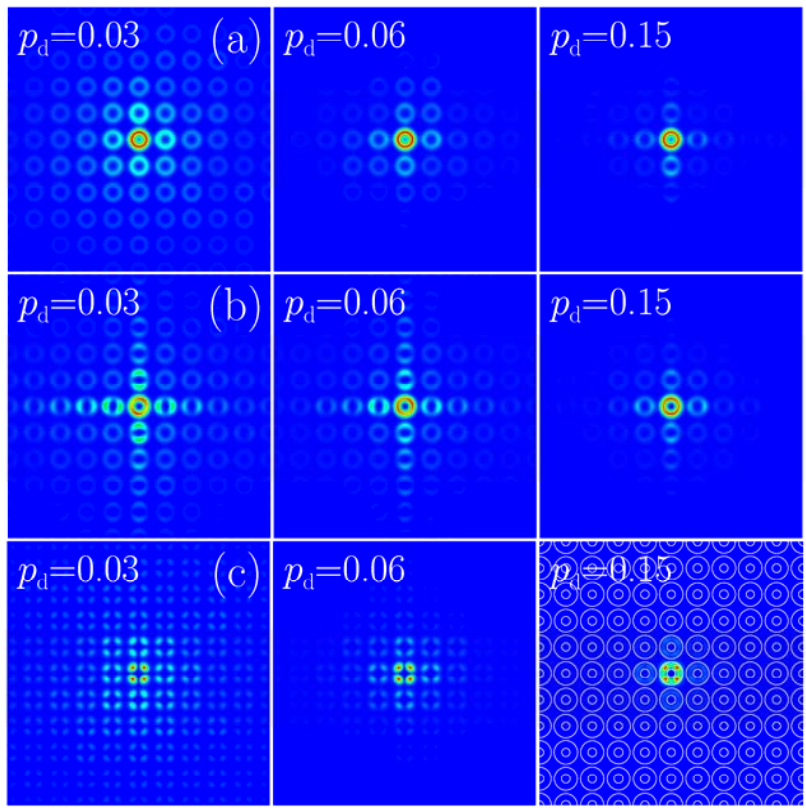

Figure 2. (Color online) Statistically averaged output intensity distributions at $\xi=1500$ for different disorder levels and input beams with $m=0$ (a), $m=1$ (b), and $m=2$ (c). White circles in the last panel in row (c) show positions of waveguides in the array. All quantities are plotted in arbitrary, dimensionless units.

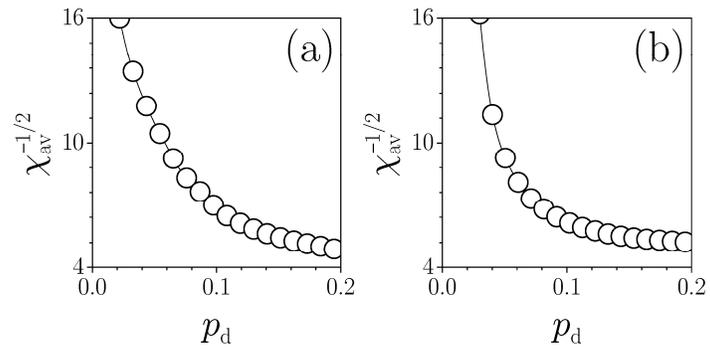

Figure 3. Averaged output inverse form-factor at $\xi=1500$ versus disorder level for the input beams with $m=1$ (a) and $m=2$ (b). All quantities are plotted in arbitrary, dimensionless units.

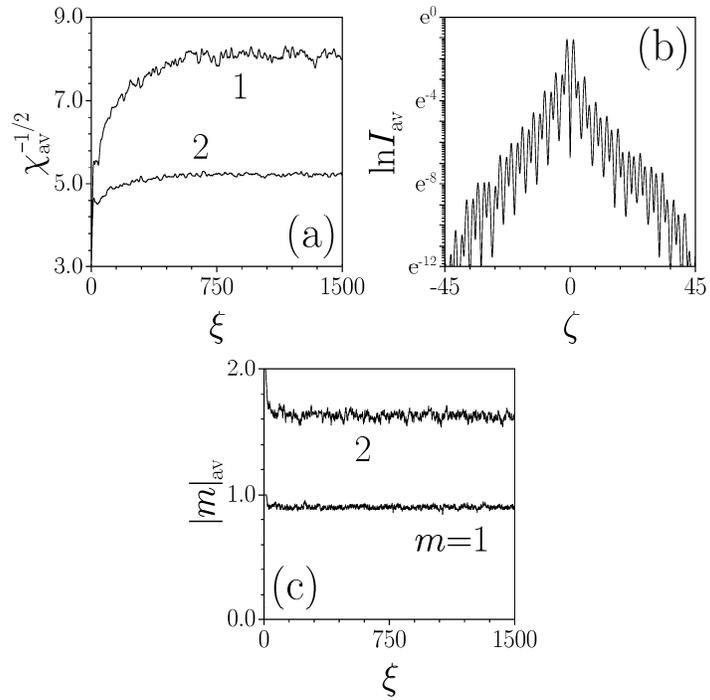

Figure 4. (a) Statistically averaged inverse form-factor versus propagation distance at $p_d = 0.08$ (curve 1) and $p_d = 0.175$ (curve 2) for the input beam with topological charge $m = 1$. (b) The cross-section of the averaged output intensity distribution at $\eta = 0$ for the input beam with $m = 1$ drawn in the logarithmic scale. The disorder level is $p_d = 0.15$. (c) Averaged modulus of topological charge of the central phase singularity versus $\xi$ at $p_d = 0.15$ and for different topological charges of the input beam. All quantities are plotted in arbitrary, dimensionless units.

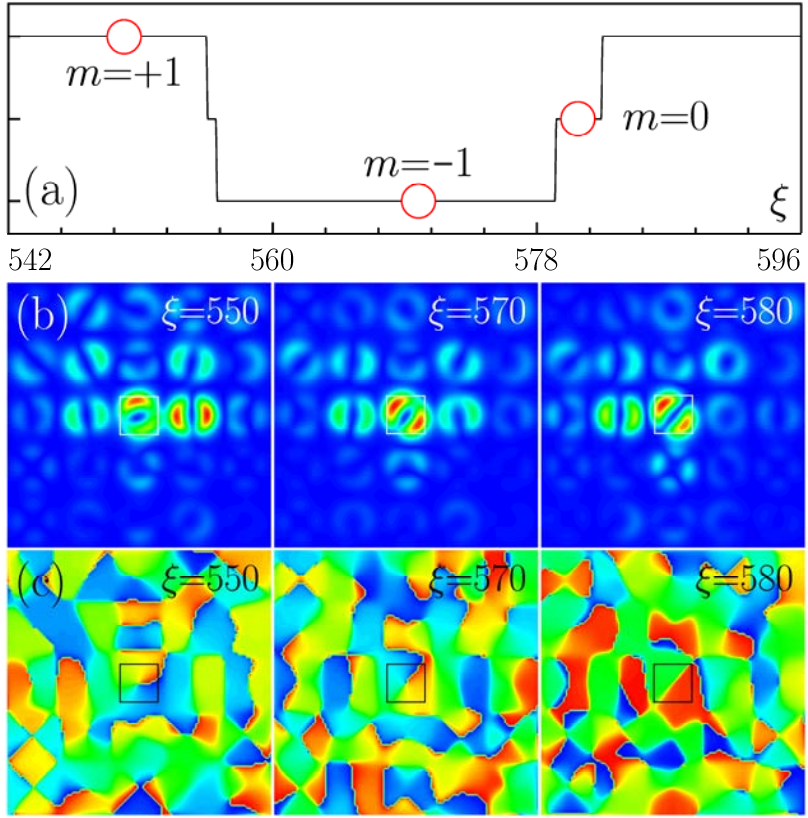

Figure 5. (Color online) (a) Topological charge of phase singularity located in the central waveguide versus propagation distance. Field modulus (b) and phase (c) distributions corresponding to the points marked by circles in panel (a). In all cases the input beam carries topological charge $m=1$ and disorder level is $p_{\rm d}=0.15$. The topological charge at a given distance $\xi$ is defined using phase circulation over closed contour indicated in (b) and (c). All quantities are plotted in arbitrary, dimensionless units.